\documentclass[preprint,aps,amsmath,nofootinbib,tightenlines]{revtex4}
\usepackage{graphicx}
\usepackage{bm}
\usepackage{epsfig}

\usepackage{subfigure}

\usepackage[hypertex]{hyperref}

\def\OMIT#1{}

\newcommand{\nn}{\nonumber} 

\newcommand{\bea}{\begin{eqnarray}}
\newcommand{\eea}{\end{eqnarray}}

\begin{document}
\setlength\baselineskip{15pt}



\preprint{ \vbox{ \hbox{arXiv:0811.3964} \hbox{CALT-68-2709} }}

\title{\phantom{x}
\vspace{0.5cm}
Trispectrum versus Bispectrum \\[6pt] 
in Single-Field Inflation
\vspace{0.5cm}
}

\author{Kevin T. Engel}
\affiliation{California Institute of 
Technology,
Pasadena, CA 91125\footnote{Electronic address: kte@caltech.edu,
ksml@caltech.edu, wise@theory.caltech.edu}
\vspace{0.2cm}}
\author{Keith S. M. Lee}
\affiliation{California Institute of 
Technology,
Pasadena, CA 91125\footnote{Electronic address: kte@caltech.edu,
ksml@caltech.edu, wise@theory.caltech.edu}
\vspace{0.2cm}}
\author{Mark B. Wise}
\affiliation{California Institute of 
Technology,
Pasadena, CA 91125\footnote{Electronic address: kte@caltech.edu,
ksml@caltech.edu, wise@theory.caltech.edu}
\vspace{0.2cm}}

\vspace{0.2cm}
\vspace{1cm}

\begin{abstract}
\vspace{0.3cm}

In the standard slow-roll inflationary cosmology, quantum fluctuations
in a single field, the inflaton, generate approximately Gaussian primordial 
density perturbations.  At present, the bispectrum and trispectrum 
of the density perturbations have not been observed and the 
probability distribution for these perturbations is 
consistent with Gaussianity. However, Planck satellite data will 
bring a new level of precision to bear on this issue, and it is possible 
that evidence for non-Gaussian effects in the primordial distribution will 
be discovered. One possibility is that a trispectrum will be observed 
without evidence for a non-zero bispectrum. 
It is not difficult
for this to occur in inflationary models where quantum fluctuations in a
field other than the inflaton contribute to the density perturbations. A
natural question to ask is whether such an observation would rule out the
standard scenarios.
We explore this issue and find that 
it is possible to construct single-field models in which inflaton-generated 
primordial density perturbations have an observable trispectrum, but a 
bispectrum that is too small to be observed by the Planck satellite.
However, an awkward fine tuning seems to be unavoidable.

\end{abstract}

\maketitle

\section{Introduction} \label{sect_intro}

Inflationary cosmology is a scenario that solves the horizon and flatness
problems \cite{Guth:1980zm, Linde:1981mu, Albrecht:1982wi}. Furthermore,
it provides a method for generating the approximately scale-invariant
primordial density perturbations that are responsible for the large-scale 
structure in the universe, as well as the anisotropy of the cosmic
microwave background radiation \cite{Mukhanov:1990me, Liddle:1993fq,
Bardeen:1983qw}. Although inflation has become the standard paradigm
for early-universe cosmology, there is no direct evidence that supports 
this paradigm, since it occurs at a very high energy scale. In the
standard slow-roll inflationary cosmology a single field, the inflaton,
is responsible for inflation and generates the primordial density
perturbations. In this case, the density perturbations  are approximately
Gaussian \cite{Maldacena:2002vr,Acquaviva:2002ud}. However, other mechanisms 
for generating the density perturbations can give rise to significant 
non-Gaussian effects \cite{Bartolo:2004if}. For example,
DBI inflation and its generalizations, in which there is a small speed
of sound, $c_s$, during inflation, give rise to  a large bispectrum,
and the present limits from WMAP already constrain such models
\cite{Silverstein:2003hf, Alishahiha:2004eh, ArmendarizPicon:1999rj,
Garriga:1999vw}. 

The Planck satellite will take us to a new level of precision in the 
measurement of the anisotropy of the microwave background radiation, 
and it is possible that non-Gaussianities will be observed 
\cite{Komatsu:2001rj, Kogo:2006kh}. 
In particle physics, it is not unusual to consider a
scalar field theory that has a connected four-point correlation but no 
three-point correlation. For example, if a $\phi \rightarrow -\phi$ 
symmetry is imposed then the three-point correlation vanishes. This is 
essentially the same mechanism that causes the model of 
Ref.~\cite{Allen:1987vq}
to have a significant trispectrum but no bispectrum.
The same signature of non-Gaussianities can also appear in the curvaton
model if some cancellation of terms occurs
\cite{Enqvist:2008gk, Sasaki:2006kq, Byrnes:2006vq}. 

A natural question that arises within the 
inflationary paradigm is whether the observation of a trispectrum for the 
primordial density perturbations but no bispectrum would 
imply that there must be more than one scalar field playing a role in 
inflation.
In this paper, we explore whether models with a single scalar field that is 
responsible for both inflation and the generation of the density perturbations 
can give rise to 
such a signature in the Planck data.
We find that this is possible, but that it seems to require a fine 
tuning of parameters. Our work shows that such an observation would not 
rule out single-field models but, because of this fine tuning, 
we would view these models as disfavored.

The rest of this paper is organized as follows. In Sec.~\ref{review},
we  briefly review the formalism used for general inflationary models.
In Sec.~\ref{nongauss}, we discuss non-Gaussianities in primordial
density perturbations. 
Some general models that feature a large trispectrum but small bispectrum 
are constructed in Sec.~\ref{models}.
Then, in Sec.~\ref{numericresults}, we consider specific examples and 
perform numerical studies of their behaviour, in particular investigating
issues of fine tuning. 
We conclude in Sec.~\ref{conclusion}.


\section{Formalism} \label{review}

Consider a general Lagrangian density for the inflaton
of the form ${\cal L} = \sqrt{-g} P(X,\phi)$, where
$X = - \frac{1}{2} g^{\mu\nu} \partial_\mu \phi \partial_\nu \phi$.
Here, $P$ plays the role of pressure.
Such a setup was studied in Ref.~\cite{Garriga:1999vw}, where it was 
applied to kinetically driven inflation (referred to as ``$k$-inflation''). 
The background solution for the inflaton is taken to be spatially
homogeneous, $\phi = \phi(t)$. Then, $X = \frac{1}{2} \dot{\phi}^2$.
$P(X,\phi)$ should satisfy the
following requirements:
\begin{align}
(i)& \qquad  \frac{\partial P}{\partial X} \geq 0 \,, 
      \label{eq:de1} \\
(ii)& \qquad  X \frac{\partial P}{\partial X} - P \geq 0 \,, \\
(iii)& \qquad  2X \frac{\partial^2 P}{\partial X^2}
                + \frac{\partial P}{\partial X} > 0 \,.
\end{align}
The first two requirements are implied by the dominant energy condition,
while the last condition (see e.g. Ref.~\cite{Bruneton:2007si})
ensures that the theory is well-defined.

The evolution of the universe is governed by the Friedmann 
and continuity equations,
\begin{align}
H^2 &= \frac{1}{3 M_P^2} \rho \,, \\
\dot{\rho} &= -3H(\rho + P) \,, \nn 
\end{align}
where $H$ is the Hubble parameter, $M_P = (8\pi G)^{-1/2}$ is the 
reduced Planck mass, and the energy density, $\rho$, is given by
\begin{align} \label{eq:rho} 
\rho &= 2 X \frac{\partial P}{\partial X} - P \,.
\end{align} 

The familiar case of slow-roll inflation utilizes a flat potential,
whose slope and curvature are characterized by the slow-roll
parameters $\epsilon\,,\,\,\eta \ll 1$. In general, however, the
potential may be relatively steep. 
An example is provided by DBI inflation, in which the inflaton $\phi$
corresponds to the position of a D3-brane rolling down a warped
throat. The Lagrangian is of the form
\begin{align}
P &= - f(\phi)^{-1} \left[\sqrt{1-2Xf(\phi)} -1 \right] - V(\phi) \,,
\end{align}
and the warping results in a speed limit of 
$X \rightarrow 1/\left(2f(\phi)\right)$.

The slow-roll parameters are then generalized to
\begin{align}
\epsilon & \equiv - \frac{\dot{H}}{H^2} 
= \frac{3 X \frac{\partial P}{\partial X}}
                 {2 X \frac{\partial P}{\partial X} - P} \,,
\label{sloweps} \\
\eta & \equiv \frac{\dot{\epsilon}}{\epsilon H} \,, 
\\
s & \equiv \frac{\dot{c}_s}{c_s H} \,, 
\end{align}
where the ``speed of sound'', $c_s$, is given by
\begin{equation} \label{eq:cs2}
 c_s^2 = \frac{\frac{\partial P}{\partial X}}
         {\frac{\partial P}{\partial X}
         + 2 X \frac{\partial^2 P}{\partial X^2}} \,.
\end{equation}
In models with a  standard kinetic term, $c_s=1$ and $s=0$.
In contrast, $c_s \ll 1$ is possible in DBI inflation; 
this leads to significant non-Gaussianities, as will be
described in the following section.

The running of the power spectrum is parameterized by the spectral
index, $n_s$, which is related to the generalized slow-roll parameters 
by
\begin{equation}
n_s - 1 = -2\epsilon - \eta - s \,.
\end{equation}
Since WMAP observes an almost scale-invariant power spectrum,
the three slow-variation parameters are constrained to be of order
$10^{-2}$.

\section{Non-Gaussianities} \label{nongauss}

It is anticipated that the Planck satellite will provide sufficiently
precise data for it to be possible to begin distinguishing between various 
inflationary models. 
One would like to be able to rule out, or at least disfavor,
broad classes of models. (A formalism for reconstructing a general
single-field action from phenomenological inputs is given in 
Ref.~\cite{Bean:2008ga}.) In this endeavour, several observables will
play a critical role, namely the spectral index, tensor perturbations
(primordial gravity waves), and non-Gaussianity. If the density 
perturbations are Gaussian, with uncorrelated Fourier modes, then the
two-point correlation function fully characterizes the distribution.
In particular, all odd correlation functions are zero, while 
higher-order even correlation functions can be expressed in terms of 
the two-point function. For example, the four-point function of a
Gaussian variable $\zeta$ is
$ \langle \zeta_1 \zeta_2 \zeta_3 \zeta_4 \rangle
= \langle \zeta_1 \zeta_2 \rangle   \langle \zeta_3 \zeta_4 \rangle 
+ \langle \zeta_1 \zeta_3 \rangle   \langle \zeta_2 \zeta_4 \rangle 
+ \langle \zeta_1 \zeta_4 \rangle   \langle \zeta_2 \zeta_3 \rangle 
$.
Thus, the detection of a three-point function or a `connected' part of 
the four-point function (or, equivalently, their Fourier transforms, 
the bispectrum and trispectrum, respectively) would signal non-Gaussianities. 
Correlations of this kind are predicted to be undetectably small
in slow-roll models of inflation, but can be large in models with a
non-standard kinetic term, such as DBI inflation.

It is convenient to characterize the size of the
bispectrum by the parameter $f_{\rm NL}$, given by
\cite{Komatsu:2001rj}
\begin{align}
\langle \zeta({\bf k}_1) \zeta({\bf k}_2) \rangle 
&= (2\pi)^3 \delta^3({\bf k}_1 + {\bf k}_2) P_\zeta(k) \,, 
\\
\langle \zeta({\bf k}_1) \zeta({\bf k}_2) \zeta({\bf k}_3) \rangle 
&= -(2\pi)^3 \delta^3({\bf k}_1 + {\bf k}_2 + {\bf k}_3)
 \frac{6}{5} f_{\rm NL} \left[ P_\zeta(k_1) P_\zeta(k_2) 
  + \hbox{perms.} \right] \,,  
\end{align}
where $\zeta$ is the gauge-invariant curvature perturbation and
``perms'' stands for two permutations of the indices.
The general expression for the  power spectrum appearing in
the formulae above is \cite{Garriga:1999vw}
\begin{equation}
 P_\zeta(k) = \frac{1}{4 M_P^2}\frac{1}{k^3} \frac{H^2}{c_s \epsilon} \,.
   \label{Pzeta}
\end{equation}
In general, $f_{\rm NL}$ will be a function of the wave numbers ${\bf k}_i$
\cite{Maldacena:2002vr}.
In practice, a comparison with data will involve evaluating
the bispectrum for a particular configuration of the ${\bf k}_i$, 
conventionally the `equilateral' configuration, in which $k_i = k$. %
Then one can set up an estimator for bispectra that have momentum 
dependences different from that of the `local' non-Gaussianity,
where $f_{\rm NL}$ is a constant \cite{Babich:2004gb, Creminelli:2006rz}.

Likewise, the size of the trispectrum is conveniently
characterized by the parameter $\tau_{\rm NL}$, which is given by
\cite{Boubekeur:2005fj}
\begin{align} \label{def:tau}
\langle \zeta({\bf k}_1) \zeta({\bf k}_2) \zeta({\bf k}_3) 
\zeta({\bf k_4}) \rangle_{\rm c}
&= (2\pi)^3 \delta^3({\bf k}_1 + {\bf k}_2 + {\bf k}_3 + {\bf k}_4)
 \frac{1}{2} \tau_{\rm NL} \left[ P_\zeta(k_1) P_\zeta(k_2) P_\zeta(k_{14})
   + \hbox{perms.}  \right]\,,
\end{align}
where the subscript ``c'' indicates the connected part,
$k_{14} = |{\bf k}_1 + {\bf k}_4|$ and 
``perms'' stands for 23 permutations of the indices.
(In fact, half of the permutations are equal to the other half,
as a consequence of the condition $\sum_i {\bf k}_i = 0$.)
In general, $\tau_{\rm NL}$ will depend upon the ${\bf k}_i$ and,
unlike $f_{\rm NL}$, even in the equilateral configuration will still 
have angular dependence (and is not restricted to lie in a plane). 
One can then choose a particular angular configuration,
by specifying the three angles $\cos \theta_{m4}$, $m=1,2,3$, 
where $\theta_{ij}$ is the angle between ${\bf k}_i$ and ${\bf k}_j$.
The current bounds from WMAP and COBE are
$-4 < f_{\rm NL}^{\rm local} < 80$ \cite{Smith:2009jr},
$-151 < f_{\rm NL}^{\rm equil} < 253$ \cite{Komatsu:2008hk},
and $|\tau_{\rm NL}^{\rm local}| \lesssim 10^8$ \cite{Boubekeur:2005fj}, 
but Planck is expected to achieve a sensitivity down to
$|f_{\rm NL}^{\rm local}| \approx 5$ \cite{Komatsu:2001rj},
$|f_{\rm NL}^{\rm equil}| \approx 66$ (at $1 \,\sigma$) 
\cite{Smith:2006ud,Bartolo:2008sg}
and $|\tau_{\rm NL}^{\rm local}| \approx 560$ \cite{Kogo:2006kh}. 

The general expression for $f_{\rm NL}$ was derived in 
Ref.~\cite{Chen:2006nt}, which built upon work in Ref.~\cite{Seery:2005wm}. 
The momentum dependence of the 
three-point function was decomposed into six functions, four of
which are suppressed by slow-roll parameters.
In the equilateral configuration, the two remaining contributions
give 
\begin{equation} \label{fNLc}
f_{\rm NL}^c = \frac{35}{108}\left(\frac{1}{c_s^2}-1 \right)\,,
\end{equation}
and
\begin{equation} \label{fNLlambda}
f_{\rm NL}^\lambda = -\frac{5}{81} \left[ \left(\frac{1}{c_s^2} - 1
   - \frac{2\lambda}{\Sigma}\right)
   + (3 - 2 {\bf c}_1) \frac{l\lambda}{\Sigma} \right] \,,
\end{equation}
where ${\bf c}_1 = 0.5772 \ldots$ is the Euler-Mascheroni constant
and
\begin{align}
  \lambda & \equiv X^2 P_{,XX} + \frac{2}{3} X^3 P_{,XXX} \,, \\
  \Sigma  & \equiv X P_{,X} + 2 X^2 P_{,XX} \,. \nn
\end{align}
Here,  $l \equiv \dot{\lambda}/(\lambda H)$ is a fourth
slow-variation parameter, and we have adopted the notation 
$P_{,X} = \partial P/ \partial X$, etc.
The formulae above make it clear that a large $f_{\rm NL}$ requires either
$c_s^2 \ll 1$ or $\lambda/\Sigma \gg 1$.  For standard slow-roll models,
$f_{\rm NL}$ is negligible, since $c_s^2 = 1$ and $\lambda=0$.  DBI inflation,
on the other hand, predicts a large bispectrum, with both $c_s^2 \ll 1$ and
$\lambda/\Sigma \gg 1$.  In this particular case, $f_{\rm NL}^c$ gives the
dominant contribution, as the first three terms in Eq.~(\ref{fNLlambda}) 
cancel, so that $f_{\rm NL}^\lambda$ is suppressed by a factor of $l$.
%

A model with an unobservable bispectrum, then, must have
$c_s^2 \approx 1$ and $\lambda/\Sigma \lesssim {\cal O}(1)$. 
In this limit, the only potentially sizeable contribution to $\tau_{\rm NL}$ 
comes from a term analogous to the $\lambda/\Sigma$ term in $f_{\rm N L}$.  
Just as $f_{\rm NL}$ can be large if $P_{,XXX} \gg 1$, $\tau_{\rm NL}$ can 
be large if $P_{,XXXX} \gg 1$. 
This leading piece can be determined from  Eqs.~(\ref{Pzeta}) and 
(\ref{def:tau}),
and the trispectrum calculations in Refs.~\cite{Huang:2006eh,Arroja:2008ga}.
We obtain
\begin{align} \label{eq:tau}
\tau_{\rm NL} \left[ \frac{1}{k_1^3}\frac{1}{k_2^3}\frac{1}{k_{14}^3}
 + \hbox{perms.} \right]
&= 384 \frac{1}{k_1 k_2 k_3 k_4} \frac{1}{K^5}
   X^4 P_{,XXXX} \frac{c_s^2}{M_P^2 H^2 \epsilon} + \cdots \,,
\end{align}
where $K = \sum_i k_i$. 
For the equilateral configuration, in which $k_i = k$
and $\sum_{m=1}^{3} \cos \theta_{m4} = -1$,
Eq.~(\ref{eq:tau}) becomes
\begin{align} \label{eq:tautheta}
\tau_{\rm NL} &= \frac{3\sqrt{2}}{32}
              \frac{1}{\sum_{m=1}^{3}(1+\cos \theta_{m4})^{-3/2}}
X^4 P_{,XXXX} \frac{c_s^2}{M_P^2 H^2 \epsilon} + \cdots \,.
\end{align}
Finally, choosing the configuration $\cos \theta_{m4} = -1/3$,
which maximizes Eq.~(\ref{eq:tautheta}),
we obtain
\begin{align} \label{eq:tau2}
\tau_{\rm NL} &= \frac{\sqrt{3}}{72} \; X^4 P_{,XXXX}
 \frac{c_s^2}{M_P^2 H^2 \epsilon}
+ \cdots \,,
\end{align}
which, combined with Eqs.~(\ref{sloweps}) and (\ref{eq:cs2}), gives
\begin{align} \label{eq:tau3}
\tau_{\rm NL}
          &= \frac{\sqrt{3}}{72} \;
             \frac{X^4 P_{,XXXX}}
                  {X P_{,X} + 2 X^2 P_{,XX}} + \cdots \,.
\end{align}
(Recall that $c_s^2 \approx 1$ in Eqs.~(\ref{eq:tau})--(\ref{eq:tau3}).)

In many inflationary models, the magnitudes of $f_{\rm NL}$ and
$\tau_{\rm NL}$ are either both small or both large. In slow-roll models,
for example, $\tau_{\rm NL} \sim f_{\rm NL} \sim \epsilon$
\cite{Seery:2006vu,Seery:2008ax}, whereas in DBI inflation we find that
(see Eq.~(\ref{fNLlambda}) and Ref.~\cite{Huang:2006eh})
\begin{equation}
\tau_{\rm NL}^{\rm DBI} = - \frac{729}{1225 \sqrt{3}} f_{\rm NL}^2 + \cdots\,.
\end{equation}

In the following section, we address the question of whether it
is possible to construct a single-field model that has a large 
trispectrum but small bispectrum.
Such a non-Gaussian signature is possible in the curvaton model,\footnote{In 
this case, the trispectrum is described by two parameters, $\tau_{\rm NL}$ 
and $g_{\rm NL}$, and it is the latter that can be very large. Of course,
consideration of a specific configuration effectively discards
such distinctions between different shapes (momentum dependences).}
provided that it has a self-interaction term in its potential.
In this scenario, the inflaton drives inflation as usual, but a
separate scalar field, known as the curvaton, produces the curvature
perturbations.

In order to have $\tau_{\rm NL} \gg f_{\rm NL}$, Eqs.~(\ref{fNLlambda}) and 
(\ref{eq:tau3}) indicate that one should have
$P_{,XXXX} \gg P_{,XXX}$. This will occur if  $\dot{\phi}$ 
converges to a speed limit at which the fourth derivative 
is singular while the third derivative is non-singular (or mildly singular).
Generically, such a model will involve a fine tuning, i.e. there
will be a relationship between the potential, $V$, and other terms in the
Lagrangian, as can be seen by simply substituting the speed limit
into the equation of motion. 
DBI inflation, however, is an exception: the speed limit occurs 
without imposing any restriction on $V$. This is because even the
first derivative has a non-analyticity; hence, in the equation of
motion, as $1/\sqrt{1-f \dot{\phi}^2} \rightarrow \infty$ the terms 
involving $V$ vanish. If, on the other hand, neither the first nor
the second derivative blows up, then there is no divergence in
the equation of motion.


\section{Model building} \label{models}
We are interested in constructing a model with $P_{,XXXX}$ parametrically
larger than $P_{,XXX}$.  One obvious solution is obtained by taking the DBI
kinetic term to a higher power:
\begin{equation}
P(X,\phi) \sim {f(\phi)}^{-1} \left(1-f(\phi) X\right)^\alpha \,.
\end{equation}
For $3 < \alpha < 4$, as $X \rightarrow f^{-1}$ the fourth derivative
becomes singular while the lower derivatives tend asymptotically to $0$.  
However, if this is the only kinetic term, the theory predicts a large 
$f_{\rm NL}$.  
Applying Eq.~(\ref{eq:cs2}) to our model, we find that in the limit 
$X \rightarrow f^{-1}$
\begin{equation}
c_s^2 \sim \frac{1-fX}{2(1-\alpha)} \rightarrow 0 \,.
\end{equation}
Because small $c_s^2$ corresponds to large $f_{\rm NL}$, this must be avoided.  
The simplest remedy is to insert a standard kinetic term.  With the
addition of functional coefficients and a potential, our first model 
(referred to henceforth as the ``$\alpha$ model'') is given by
\begin{equation}
P(X,\phi) = AX+Bf^{-1}\left(1-fX\right)^\alpha-U \, ,
\end{equation}
where $A$, $B$, $f$, and $U$ can all be functions of $\phi$, and $U = V + Bf^{-1}$.  A few restrictions can be imposed on this theory from the 
beginning.  First, for small $X$, the second term can be expanded, generating
the standard canonical kinetic term if we require
\begin{equation}
\left(A-\alpha B\right)|_{\phi = 0} = 1 \,.
\end{equation}
Another constraint is given by the energy density, which can be calculated from
Eq.~(\ref{eq:rho}):
\begin{equation}
\rho = AX-B\left(1-fX\right)^{\alpha-1}\left[\left(2\alpha-1\right)X+f^{-1}
\right] + U \,.
\end{equation}
Unlike DBI inflation, this theory contains no singular terms in its equation
of motion; hence $X$ could potentially surpass $f^{-1}$.  In order
to avoid imaginary terms in the energy density, $\alpha$ must therefore have
an odd denominator.

With these restrictions in place, a given set of parameters produces a
sensible theory where $\phi$ rolls to the minimum of the potential.
$\tau_{\rm NL}$ becomes large, however, only when $\frac{1}{2}\dot{\phi}^2$ 
is close to $f^{-1}$.  In order to produce large non-Gaussianities 
from generic initial conditions, we must therefore impose a speed limit.
We do this by inserting our desired speed limit into the equation of 
motion for $\phi$:
\begin{equation}
X = \frac{1}{2}\dot{\phi}^2 \rightarrow \frac{1}{f}, \quad \ddot{\phi} 
\rightarrow \frac{-f^\prime}{f^2} \, ,
\end{equation}
where the prime denotes a derivative with respect to $\phi$.  With these 
insertions several terms drop out, and the remainder can be 
integrated to give 
\begin{equation}\label{V}
\left(U+\frac{A}{f}\right)^{1/2} = \frac{\sqrt{6}}{2 M_P}
\int \frac{A}{\sqrt{f}}\, d\phi \,.
\end{equation}
To obtain a speed limit, we are forced to have this relationship between
the kinetic and potential terms.  For given $A$, $f$, and $B$, $V$ is 
nearly completely specified (some minimal freedom in lower-order terms 
exists due to the constant of integration).

We can check the stability of this solution by expanding around the speed 
limit, writing
$X \rightarrow \frac{1}{f}(1-\varepsilon)$.  For small $\varepsilon$, we
find that
\begin{equation} \label{epsdot}
\frac{\dot{\varepsilon}}{\varepsilon}
= \sqrt{\frac{2}{f}}\left[\frac{A^{\prime}}{A} - \frac{f^{\prime}}{f}\right]
   - \frac{\sqrt{3}}{M_P}
     \left[ \left(\frac{A}{f}+U \right)^{1/2} + 
            \frac{A}{f(\frac{A}{f}+U)^{1/2}}\right] \,.
\end{equation}
For the models we consider, the negative second term is dominant, thereby 
driving $\varepsilon$ to $0$.  With $\dot{\phi}$ forced to the speed limit, 
the non-Gaussianities are straightforward to calculate.  In the equilateral
configuration, the dominant contribution to $\tau_{\rm NL}$ is given by
\begin{equation}\label{tau1}
\tau_{\rm NL} = \frac{\sqrt{3}}{72}\frac{B}{A} \alpha(\alpha-1)(\alpha-2)
(\alpha-3)\varepsilon^{\alpha-4} \,.
\end{equation}
The main contribution to $f_{\rm NL}$, on the other hand, is given by
\begin{equation}
f_{\rm NL} \sim \frac{B}{A} \varepsilon^{\alpha-3} \,.
\end{equation}
As the inflaton rolls to its speed limit, a suitable choice for $\alpha$
renders $f_{\rm NL}$ small and $\tau_{\rm NL}$ large.  We have achieved our 
goal, encoding the non-Gaussianities for this model in the trispectrum rather
than the bispectrum, but to obtain this result we have had to tune the 
potential.  We can examine the degree of fine tuning required by perturbing 
the potential.  Making the replacements
\begin{equation}
X \rightarrow \frac{1}{f}(1-\varepsilon), \quad V \rightarrow V + \delta V \, ,
\end{equation}
we expand to first order in $\varepsilon$ and $\delta V$ and ignore 
$\dot{\varepsilon}$ terms.  The result is an equation for $\varepsilon$:
\begin{equation}\label{vareps}
  \varepsilon = \frac{f \delta V 
    - \frac{2 M_P}{\sqrt{6}} \frac{(A + f U)^{1/2}}{A} f \delta V^\prime}
       {2A + f U + \frac{2 M_P}{\sqrt{6}} (A + f U)^{1/2}
        \left( \frac{f^\prime}{f}-\frac{A^\prime}{A} \right)} \,.
\end{equation}
For unperturbed $V$, Eq.~\eqref{epsdot} shows that $\varepsilon$ decreases 
roughly exponentially.  With perturbations, however, $\varepsilon$ tends to 
level off at some non-zero value.  We shall show with specific examples in the 
next section that for large enough perturbations $\varepsilon$ does not 
become small enough to generate an observable $\tau_{\rm NL}$.  

From a tuning standpoint, 
then, we are interested in enhancing $\tau_{\rm NL}$ so that $\varepsilon$
does not have to be extremely small, thus allowing a wider range of
potentials.  
From Eq.~\eqref{tau1}, we see that a large value for $B/A$
appears to accomplish this goal. However, for generic initial 
conditions not at the speed limit, the ratio is bounded.  
The first dominant energy condition (Eq.~\eqref{eq:de1}) evaluated at 
$\dot{\phi} = 0$ gives
\begin{equation}
A-\alpha B \geq 0 \,.
\end{equation}
The same equation evaluated at the speed limit 
implies that $A$ must be positive; therefore,
for positive $B$, $B/A \leq 1/\alpha$.  For negative $B$, the restriction
appears in the equation of motion.  For large $|B/A|$ the coefficient
of $\ddot{\phi}$ can evolve to $0$, causing the equation to become
singular, unless
\begin{equation}
\left|\frac{B}{A}\right| < \frac{1}{2\alpha} 
                  \left(1-\frac{3}{2\alpha-1}\right)^{2-\alpha} \,.
\end{equation}
These two conditions force us to impose
\begin{equation}
\qquad \qquad \qquad \qquad \qquad \qquad \qquad ~~
\left|\frac{B}{A}\right|
\lesssim \frac{1}{\alpha} \qquad \mbox{(before hitting speed limit)}\,. 
\nonumber
\end{equation}
Once $\phi$ is at the speed limit, however, $B$ vanishes from the equation 
of motion and can be set arbitrarily large.  For clever choices of the 
functions $A$ and $B$, $B/A$ can be made small initially, but larger later 
on, thus enhancing $\tau_{\rm NL}$.  Despite its apparent promise, this idea
actually has limited utility, for a couple of reasons.
First, the requirement of a canonical kinetic term forces the use of 
unappealing functional forms: the simplest model we found had $B$ as a 
Gaussian in $\phi$.  More importantly, $f_{\rm NL}$ also scales with 
$B/A$, which limits the overall use of the ratio to magnify
$\tau_{\rm NL}$.  

With the constraint that $B/A$ is of order unity, the last way to enhance
$\tau_{\rm NL}$ is to choose an $\alpha$ as small as possible.  This
leads to a preferred $\alpha$ for each of the allowed odd denominators:
\begin{equation}
\alpha = \frac{10}{3},\,\frac{16}{5},\,\frac{22}{7}, \hdots \,.
\end{equation}
Considering the $\varepsilon$ dependence of $\tau_{\rm NL}$ at the
end-point of this series leads us to our second model,
designated the ``log model'':
\begin{equation}
P(X,\phi) = AX+Bf^{-1}\left(1-fX\right)^3 \log |1-fX|-U\, ,
\end{equation}
where in this model $U = V$.  Most of the equations from 
the previous model carry over.  
In particular, the potential equation \eqref{V}, the stability equation
\eqref{epsdot}, and the perturbation equation \eqref{vareps} all
continue to hold for the log model.
The canonical constraint changes to 
\begin{equation}
\left(A-B\right)|_{\phi = 0} = 1\,,
\end{equation}
and a similar analysis to the one above shows that
\begin{equation}
\qquad \qquad \qquad \qquad \qquad \qquad \qquad ~~
\left|\frac{B}{A}\right|
\lesssim 1 \qquad \mbox{(before hitting speed limit)}\,. \nonumber
\end{equation}
The main difference between the two models is the signature of
the non-Gaussianities.
In the equilateral configuration, to leading order we have
\begin{equation}
f_{\rm NL} = -\frac{40}{81} \frac{B}{A} \log |\varepsilon|\, ,
\end{equation}
\begin{equation}
\tau_{\rm NL} = \frac{\sqrt{3}}{12}\frac{B}{A}\varepsilon^{-1}\,.
\end{equation}
We see that $f_{\rm NL}$ now diverges as $\varepsilon \rightarrow 0$, albeit
more slowly than $\tau_{\rm NL}$.  Also, for the same $\varepsilon$, 
$\tau_{\rm NL}$
is larger in the log model than in the $\alpha$ model, which in turn allows 
a less fine-tuned potential.  We shall explore these models and the
requisite fine tuning with numerical simulation of specific examples
in the next section.

\section{Numerical studies} \label{numericresults}
We consider numerical solutions to the equation of motion using initial 
conditions $\phi(0) = \phi_0$ and $\dot{\phi}(0)=0$.  An initial 
velocity for $\phi$ makes little difference to the end result as $\phi$
is generally driven quickly to the speed limit.  The initial value for 
$\phi$ is set by the need for a nearly scale-invariant spectrum.  The 
slow-variation parameter $\epsilon$ is required to be 
${\mathcal O}(10^{-2})$, and this usually imposes a lower bound on $\phi$.  
Two other observational constraints that need to be considered are
the amplitude of the power spectrum and the number of e-foldings.  For 
single-field inflation the size of the density perturbations is 
characterized by
\begin{equation}\label{delrho}
\Delta_{\cal R}^2=\frac{k^3 P_\zeta}{2\pi^2}=\frac{H^2}{8\pi^2M_P^2c_s\epsilon}\, ,
\end{equation}
whose value has been observed to be approximately $2.45\times10^{-9}$ 
\cite{Komatsu:2008hk}.  The effect of this constraint in our models is to set a
mass scale.  The relevant equation for the number of e-foldings is
\begin{equation}
N_e = \int_{\phi_i}^{\phi_f} \frac{H}{\dot{\phi}}\,d\phi \,.
\end{equation}
Cosmological scales of interest exit the horizon $50-60$ e-folds before the 
end of inflation.  We need to ensure that there is a sufficiently long
observation window, during which $\tau_{\rm NL}$ should be large,  
and also that roughly 60 e-folds can be attained in total.  
Like $\epsilon$, this puts a lower bound on $\phi_0$.
Finally, one other observable of interest is $r$, the tensor-to-scalar
ratio.  Garriga and Mukhanov derived in Ref. \cite{Garriga:1999vw} 
the result that for general single-field models of inflation
\begin{equation}
r = 16 c_s \epsilon \,.
\end{equation}
For the models we consider here, $c_s = 1$.  To obtain $n_s$ consistent with
observations, $\epsilon \sim .01$; therefore, our models predict a value
for $r$ in the range $.1-.2$.  
\subsection{$\alpha$ model}
The simplest model is the one in which $A$, $B$, and $f$ are all constant:
for example,
\begin{equation}
A = 2, \quad B = \frac{1}{\alpha}, \quad 
f = \frac{6}{m^4}, \quad \alpha = \frac{10}{3} \,.
\end{equation}
The potential is then given by Eq.~\eqref{V}:
\begin{equation}
V = m^4 \left[\Bigl(\frac{\phi}{M_P}\Bigr)^2-\frac{1}{3}-
\frac{1}{6 \alpha}\right] \,.
\end{equation}
The negative energy density that can result for small $\phi$ is not a 
concern here, for it turns out that inflation occurs only for
$\phi > M_P$.  The inflation era ends when $\ddot{a} \leq 0$ or
equivalently $\epsilon \geq 1$.  Eq.~(\ref{sloweps}) evaluated at the speed
limit gives
\begin{equation}
\epsilon = \frac{M_P^2}{\phi^2} \,.
\end{equation}
We see that $\phi$ must be larger than $1~M_P$ during inflation, and 
that $\phi \sim 10\,M_P$ as current cosmological scales were exiting 
the horizon.  The number of e-folds, calculated to be
\begin{equation}
N_e = \frac{1}{2}\left[\Bigl(\frac{\phi_i}{M_P}\Bigr)^2-
\Bigl(\frac{\phi_f}{M_P}\Bigr)^2\right] \, ,
\end{equation}
imposes a similar restriction on $\phi$.  In order to achieve the final
60 e-folds, the end of the observation window must be at or beyond
$\phi = 11\,M_P$.  Finally, we can calculate $\Delta_{\cal R}^2$
from Eq.~\eqref{delrho}:
\begin{equation}
\Delta_{\cal R}^2=\frac{1}{24\pi^2}\Bigl(\frac{m}{M_P}\Bigr)^4\Bigl(\frac{\phi}{M_P}\Bigr)^4=2.45\times10^{-9}\,.
\end{equation}
Substituting in $\phi = 12\,M_P$, we find that $m \sim 2\times10^{-3}\,M_P$.  

\begin{figure}[h!] 
\subfigure[]{
\includegraphics[width=2.40in]{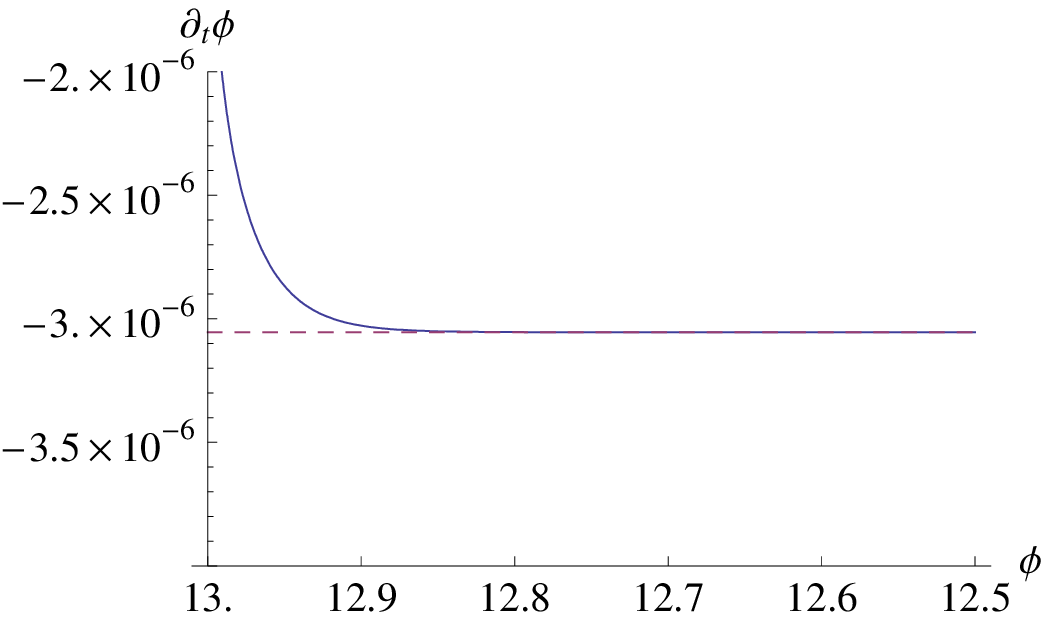}
\label{vel1}
} 
\subfigure[]{
\includegraphics[width=2.40in]{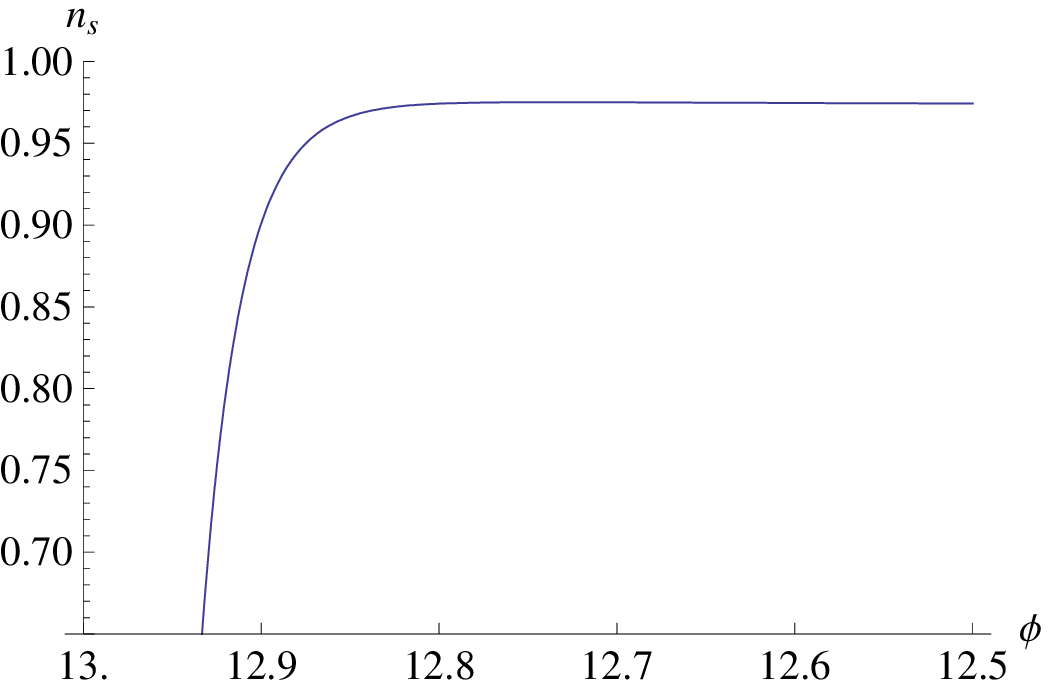}
\label{ns1}
}
\subfigure[]{
\includegraphics[width=2.40in]{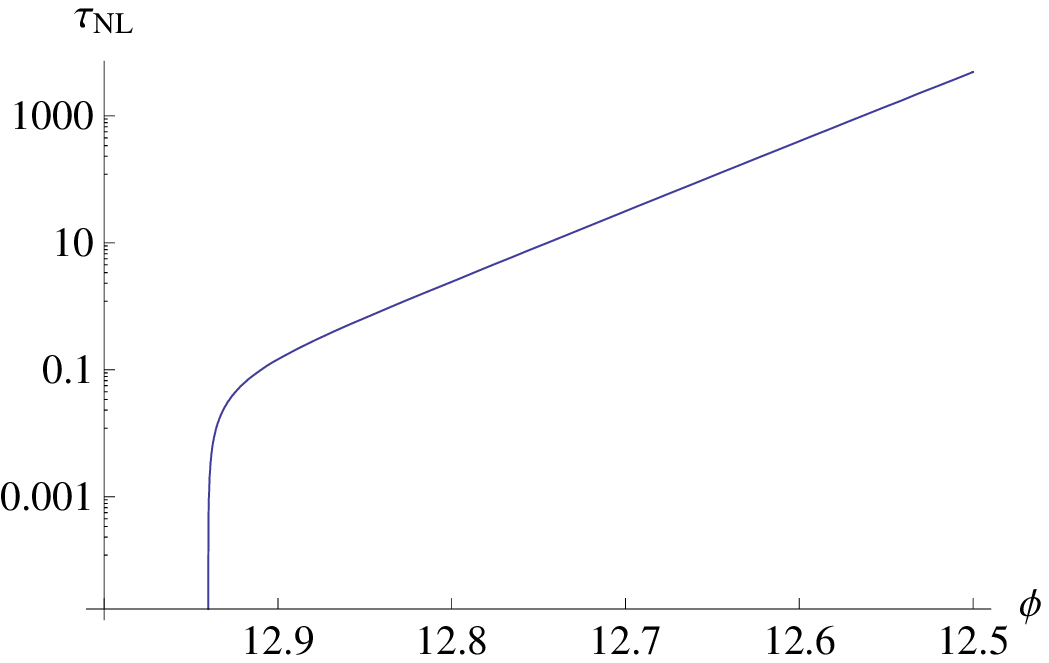}
\label{taunl1}
}
\caption{\subref{vel1} $\dot{\phi}$ (solid) plotted with speed limit of
$-\sqrt{\frac{2}{f}}$ (dashed). \subref{ns1} Spectral index $n_s$.
\subref{taunl1} $\tau_{\rm NL}$.}
\label{alpharesults}
\end{figure}

For the initial conditions $\phi(0) = 13\,M_P$, and $\dot{\phi}(0) = 0$, the
key features of the evolution are shown in Figure~\ref{alpharesults}.
As predicted, $\tau_{\rm NL}$ grows roughly exponentially
for $V$ that satisfies Eq.~\eqref{V} exactly.  Since $|\tau_{\rm NL}|$ has been
observationally constrained to be less than $10^8$, this point signifies the 
end of our possible observational window.  To enforce this, we require
some exit mechanism to kick in around $\phi = 5\,M_P$, ending inflation.
The beginning of our observational window is set by $n_s$, which becomes
flat only as $\dot{\phi}$ approaches the speed limit. For this model,
the observational window occurs for $\phi$ in the range 
$12.85\,M_P > \phi > 12.10\,M_P$, 
lasting about 9 e-folds.  
This window corresponds to the range of observable wave numbers, $k$. 
$f_{\rm NL}$ is unobservably small throughout, but $\tau_{\rm NL}$ is very 
$k$-dependent, growing from $\sim 1$ at large scales to $10^8$ at the 
smallest scales.

\begin{figure}[h!]
\includegraphics[width = 3.20in]{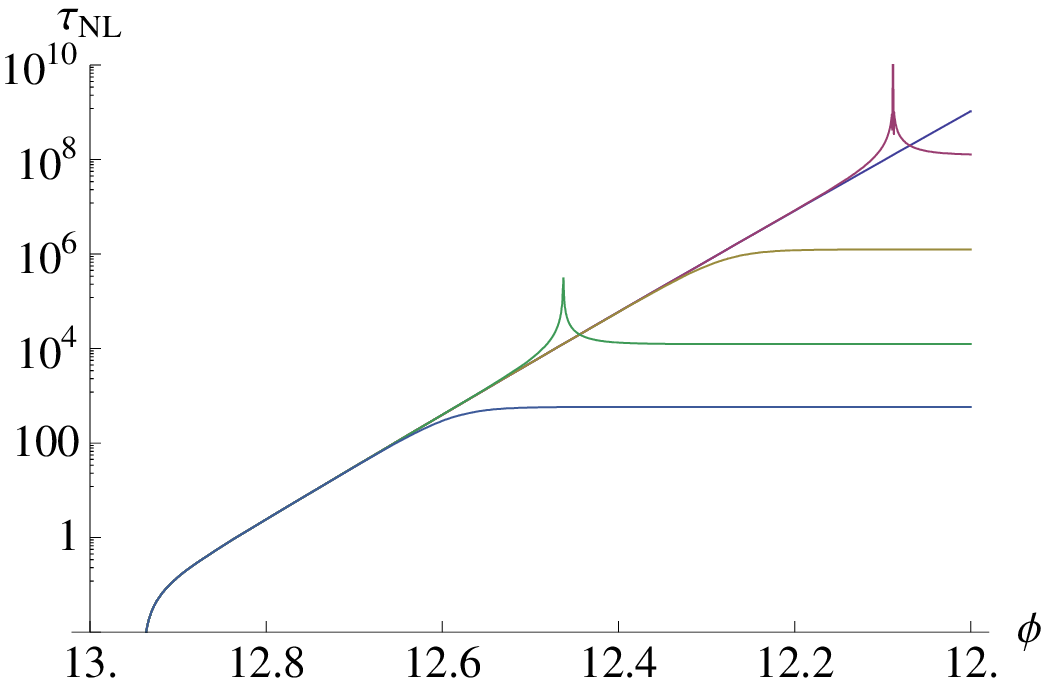}
\caption{$\tau_{\rm NL}$ plotted for various values of $\delta$.  From top 
to bottom, 
$\delta = 0\, , \, 10^{-15} \, , \, -10^{-12} \, , \, 10^{-9}
\, , \, -10^{-7}$.  
For positive $\delta$, $\dot{\phi}$ crosses the speed
limit before leveling off, creating the spikes shown in $\tau_{\rm NL}$.}
\label{vartaunl}
\end{figure}

These results are not typical ones, however: an inflaton potential
even slightly modified from one satisfying Eq.~\eqref{V} exactly 
will lead to a significantly different signature.
We investigate the allowed sizes of these deviations and their effects 
on $\tau_{\rm NL}$ by perturbing the coefficients in the potential.  
Figure~\ref{vartaunl} shows the result of 
perturbing the mass term by
\begin{equation}
\delta V = m^4 \delta \Bigl(\frac{\phi}{M_P}\Bigr)^2
\end{equation}
for several different values of $\delta$.
Altering other coefficients gives similar results; therefore, the general 
effect of potential perturbations is to cause $\tau_{\rm NL}$ to level off. 
A large enough perturbation (e.g. $|\delta_{\rm min}| \sim
10^{-15}$ for the mass term)
allows more freedom in choosing the end of the observation window since 
$\tau_{\rm NL}$ no longer saturates the $10^8$ bound.  However, we are limited 
in the size of the perturbation.  In this model, tuning the $\phi^2$ term 
by more than $10^{-7}$ suppresses the non-Gaussianities beyond the Planck
satellite's resolution.  
Table~\ref{alphatable} gives the maximum deviation allowed for each coefficient
in the potential.  In order to produce an observable
$\tau_{\rm NL}$ for this model, we conclude that the potential must be 
fine-tuned to one part in $10^7$.
\begin{table}[h!]
\begin{tabular}{|c|c|}
\hline
Coefficient&~$\delta_{\rm max}$~\\
\hline
$\phi^2$&~$10^{-7}$~ \\
$\phi^0$&~$10^{-4}$~ \\
\hline
\end{tabular}
\caption{Coefficient tuning in $V$ (i.e. $c\phi^n \rightarrow
c(1+\delta)\phi^n$) for which $|\tau_{\rm NL}|$ levels off at $\sim 500$.}
\label{alphatable}
\end{table}
\subsection{Log model}
The parameter choices made in the above example are by no means unique.
To illustrate the diversity of possible models, we switch to the log model
and consider a $\phi$-dependent speed limit:
\begin{equation}
A = 2, \quad B = 1, \quad 
f = \frac{6}{m^4}\left(\frac{M_P}{\phi}\right)^2 \,.
\end{equation}
In this case, the constant of integration occurring in Eq.~\eqref{V} is not 
just a redefinition of $\phi$ and gives some freedom in defining $V$.
We shall choose
\begin{equation}
V = m^4 \left[\frac{1}{4}\Bigl(\frac{\phi}{M_P}\Bigr)^4 +
\frac{2}{3}\Bigl(\frac{\phi}{M_P}\Bigr)^2 + 1\right] \,.
\end{equation}
For this model,
\begin{eqnarray}
\epsilon &=& \frac{\bigl(\frac{\phi}{M_P}\bigr)^2}{\left[\frac{1}{2}
\bigl(\frac{\phi}{M_P}\bigr)^2+1\right]^2} \, , \\
\Delta_{\cal R}^2 &=& \frac{1}{24\pi^2}\Bigl(\frac{m}{M_P}\Bigr)^4
\left[\frac{1}{2}\Bigl(\frac{\phi}{M_P}\Bigr)^2+1\right]^4\Bigl(\frac{M_P}{\phi}
\Bigr)^2 = 2.45\times10^{-9} \, , \\
N_e &=& \frac{1}{4}\left[\Bigl(\frac{\phi_i}{M_P}\Bigr)^2-
\Bigl(\frac{\phi_f}{M_P}\Bigr)^2\right]+\log \Bigl(\frac{\phi_i}{\phi_f}\Bigr)\,.
\end{eqnarray}
Our constraints set $\phi_0 \sim 20\,M_P$ and $m \sim 6\times10^{-4}\,M_P$.  

\begin{figure}[h!] 
\subfigure[]{
\includegraphics[width=2.4in]{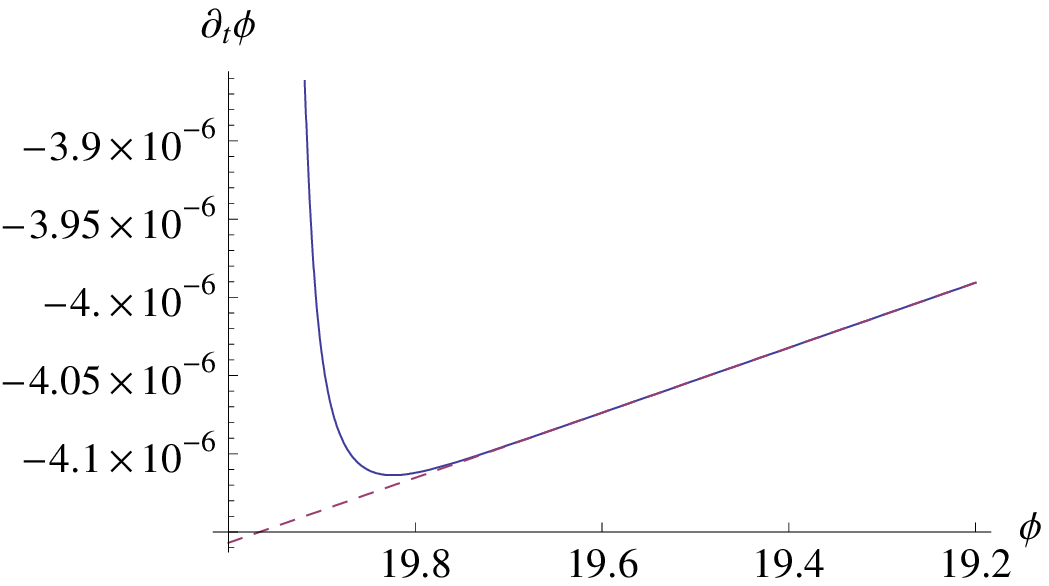}
\label{vel2}
} 
\subfigure[]{
\includegraphics[width=2.40in]{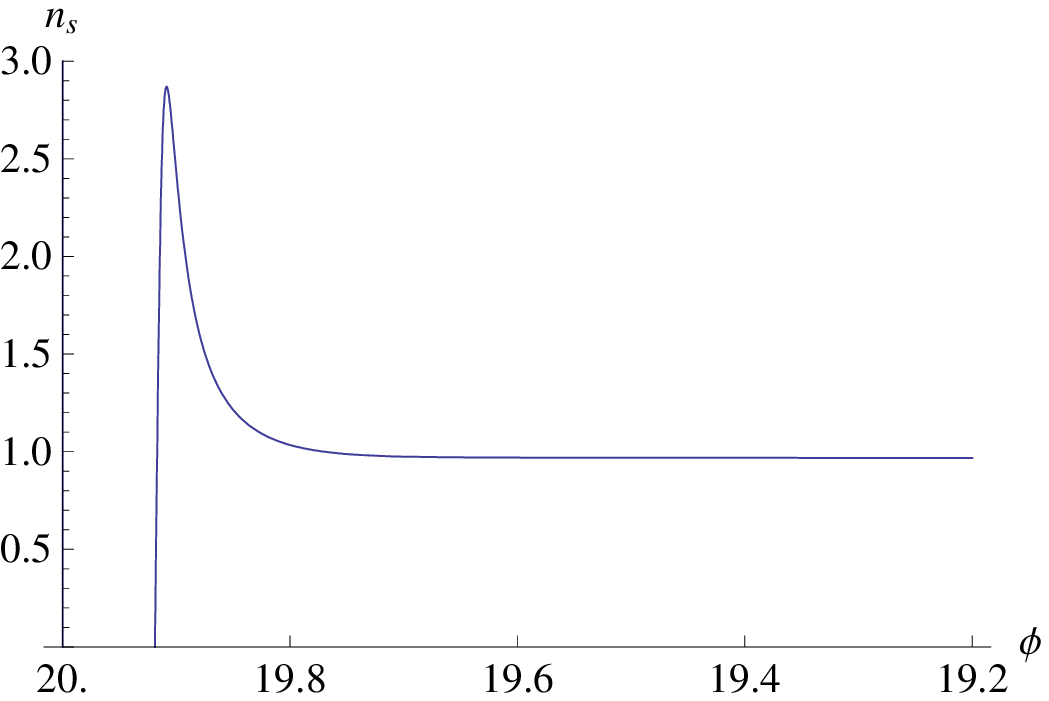}
\label{ns2}
}
\subfigure[]{
\includegraphics[width=2.40in]{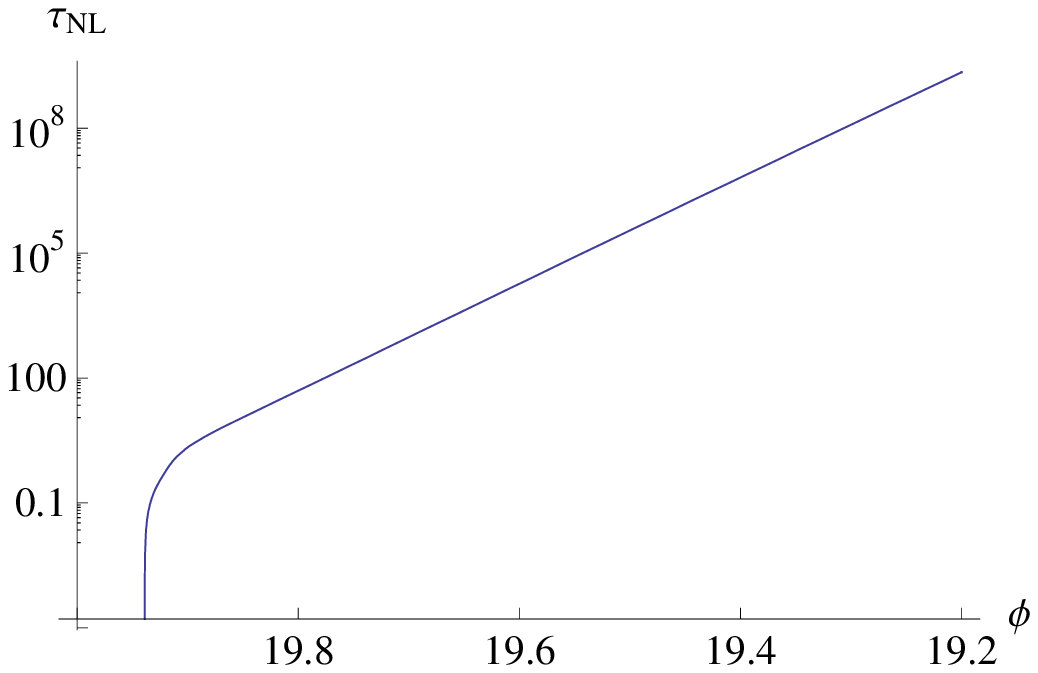}
\label{taunl2}
}
\subfigure[]{
\includegraphics[width=2.40in]{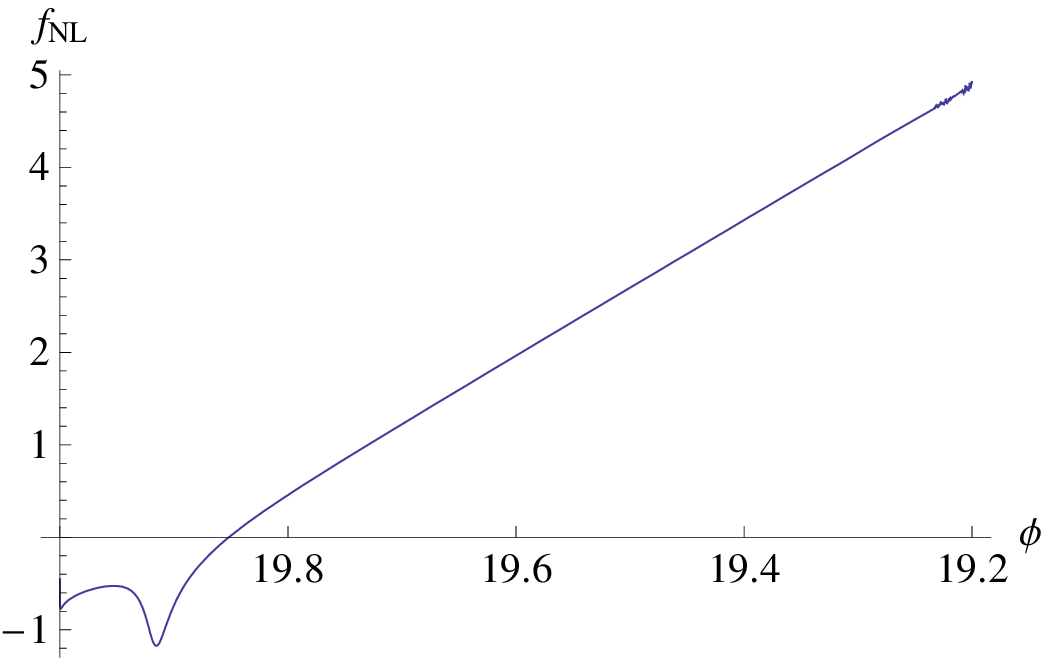}
\label{fnl1}
}
\caption{\subref{vel1} $\dot{\phi}$ (solid) plotted with speed limit of
$-\sqrt{\frac{2}{f}}$ (dashed). \subref{ns1} Spectral index $n_s$.
\subref{taunl1} $\tau_{\rm NL}$. \subref{fnl1} $f_{\rm NL}$.}
\label{logresults}
\end{figure}

For $\phi(0) = 20\,M_P$ and $\dot{\phi}(0) = 0$, the results are shown 
in Figure~\ref{logresults}.
We see the main differences from the previous example are a speed limit
that depends on $\phi$ and, since we are using the log model, an $f_{\rm NL}$
that slowly diverges as $\varepsilon \rightarrow 0$.  Even so, $f_{\rm NL}$ 
is still predicted to be unobservably small: for $\tau_{\rm NL}$ at its 
bound of $10^8$, $f_{\rm NL} \sim 5$.  For exact $V$, $\tau_{\rm NL}$ grows 
even faster than before; in this model, the possible observation
window is only about 4 e-folds in duration.  To obtain an acceptable
theory, we again consider perturbations to $V$.
Sufficiently large perturbations fix the problem, causing
$\tau_{\rm NL}$ to level off below $10^8$ ($|\delta_{\rm min}| \sim
10^{-9}$ for the quartic term).  The maximum deviation that 
still produces an observable $\tau_{\rm NL}$ is shown for each coefficient
of $V$ in Table~\ref{logtable}.  As predicted at the end of Sec.~\ref{models},
the log model is less restrictive: the potential must be fine-tuned only to 
one part in $10^4$.
\begin{table}[h!]
\begin{tabular}{|c|c|c|}
\hline
Coefficient&~$\delta_{\rm max}$~\\
\hline
$\phi^4$&~$10^{-4}$~\\
$\phi^2$&~$1$~\\
$\phi^0$&~$10$~\\
\hline
\end{tabular}
\caption{Coefficient tuning in $V$ (i.e. $c\phi^n \rightarrow
c(1+\delta)\phi^n$) for which $|\tau_{\rm NL}|$ levels off at $\sim 500$.}
\label{logtable}
\end{table}

\section{Conclusion} \label{conclusion}
We have constructed several single-field models of inflation that satisfy
current observational constraints and produce
a large trispectrum and a small bispectrum.  However, these features come
at a cost.  To generate the desired non-Gaussianities, we are forced to 
include an unmotivated non-standard kinetic term in the Lagrangian.  
The potential cannot be arbitrary, thus introducing a degree of fine tuning 
into the theory.
Of the models considered, the best cases allow a tuning of order $10^{-4}$.
Larger modifications result in a trispectrum unobservable by the Planck
satellite.  
These two features - the strange kinetic term and the fine-tuned potential -
seem to be general requirements for a single-field model predicting large
$\tau_{\rm NL}$ and small $f_{\rm NL}$.  If Planck does indeed observe a 
trispectrum but no bispectrum, we conclude that these undesirable attributes 
tend to disfavor single-field inflation.  More consideration should instead be
given to multiple-field models such as the curvaton mechanism, which can more 
naturally produce these kinds of non-Gaussianities.


\acknowledgments 
We thank Gary Shiu for correspondence regarding Ref.~\cite{Huang:2006eh}.
This work was supported in part by the U.S. Department of Energy
(DOE) under the cooperative research agreement DE-FG02-92ER40701. 
K.L. was also supported by the Sherman Fairchild Foundation.


\end{document}